\documentclass[twocolumn]{autart}    

\usepackage{graphicx}          

\usepackage{amsmath}
\usepackage{amssymb}
\theoremstyle{definition}
\newtheorem{definition}{Definition}
\theoremstyle{assumption}
\newtheorem{assumption}{Assumption}
\theoremstyle{problem}
\newtheorem{problem}{Problem}
\theoremstyle{lemma}
\newtheorem{lemma}{Lemma}
\theoremstyle{remark}
\newtheorem{remark}{Remark}
\theoremstyle{theorem}
\newtheorem{theorem}{Theorem}
\usepackage{url}
\usepackage{mathtools} 
\usepackage{breqn}

\usepackage[flushleft]{threeparttable}
\usepackage{array,booktabs,makecell}

\usepackage{xcolor}
\DeclareMathOperator*{\argmin}{arg\,min \,\,}

\newcommand{\norm}[1]{\left \lVert #1 \right \rVert}
\newcommand{\abs}[1]{\left | #1 \right |}

\begin{document}

\begin{frontmatter}

\title{Direct Data-Driven Discounted Infinite Horizon Linear Quadratic Regulator with Robustness Guarantees\thanksref{footnoteinfo}} 

\thanks[footnoteinfo]{This paper was not presented at any IFAC 
meeting. Corresponding author Hamidreza Modares. Tel. +XXXIX-VI-mmmxxi. 
Fax +XXXIX-VI-mmmxxv.}

\author[a]{Ramin Esmzad }\ead{esmzadra@msu.edu},    
\author[a]{Hamidreza Modares}\ead{modaresh@msu.edu},               

\address[a]{Department of Mechanical Engineering, Michigan State University, USA}  

\begin{keyword}                           
Direct data-driven control, linear quadratic regulator, robustness, suboptimality gap.               
\end{keyword}                             

\begin{abstract}                          
This paper presents a one-shot learning approach with performance and robustness guarantees for the linear quadratic regulator (LQR) control of stochastic linear systems. Even though data-based LQR control has been widely considered, existing results suffer either from data hungriness due to the inherently iterative nature of the optimization formulation (e.g., value learning or policy gradient reinforcement learning algorithms) or from a lack of robustness guarantees in one-shot non-iterative algorithms. To avoid data hungriness while ensuing robustness guarantees, an adaptive dynamic programming formalization of the LQR is presented that relies on solving a Bellman inequality. The control gain and the value function are directly learned by using a control-oriented approach that characterizes the closed-loop system using data and a decision variable from which the control is obtained. This closed-loop characterization is noise-dependent. The effect of the closed-loop system noise on the Bellman inequality is considered to ensure both robust stability and suboptimal performance despite ignoring the measurement noise. To ensure robust stability, it is shown that this system characterization leads to a closed-loop system with multiplicative and additive noise, enabling the application of distributional robust control techniques. The analysis of the suboptimality gap reveals that robustness can be achieved by construction without the need for regularization or parameter tuning. The simulation results on the active car suspension problem demonstrate the superiority of the proposed method in terms of robustness and performance gap compared to existing methods.
\end{abstract}

\end{frontmatter}

\section{Introduction}
Data-driven control consists of a broad family of control systems that utilize offline or real-time data to design controllers to achieve the required system specifications ~\cite{hou2013model,markovsky2021behavioral}. It departs from classical control theory, which often relies on mathematical models of the system to derive control strategies. In general, data-driven control methods can be categorized into two main classes, i.e., indirect vs. direct methods~\cite{9683187,DEPERSIS2021109548}. In indirect data-based control, first, a model of the system is identified (system identification step), and then, a controller is designed for the identified model, assuming that this model is the ground truth of the actual system. This approach to control is also called the "certainty equivalence (CE) principle." CE might not provide stability and robustness guarantees for systems under noise or disturbances~\cite{NEURIPS2019_5dbc8390}. Therefore, a robust counterpart of this indirect data-based control characterizes a set of all possible open-loop systems that conform with data (instead of identifying a point-based system model) and design a controller against the worst-case system realization inside this set~\cite{pmlr-v120-coppens20a,treven2021learning}. On the other hand, in direct data-based control, which is also called the "model-free approach," a controller is designed directly based on collected data, bypassing the system identification step. 
A comparison of direct and indirect data-driven control, along with their advantages and disadvantages, is provided in ~\cite{10061542} and \cite{NEURIPS2019_5dbc8390}.

Both direct and indirect data-driven control design methods have been widely used recently to solve different control problems, including safe control design \cite{10318172}, stable control design \cite{de2019formulas}, and optimal control design \cite{van2020noisy,gravell2019learning,NEURIPS2019_5dbc8390}, {specifically linear quadratic regulator (LQR), which is the focus of this paper. LQR aims to design a linear controller for linear systems by optimizing a quadratic cost function that penalizes both state deviations and control effort. Data-based LQR algorithms leverage the collected data and formalize the LQR control design problem as a data-dependent convex optimization problem from which a control solution is learned.} Among the direct learning of LQR control solutions, {iterative approaches such as value learning (VL) (which includes policy iteration and value iteration)\cite{cui2023lyapunov,8169685,pmlrv211cui23c,bian2016value,cui2023reinforcement}}, Q-learning~\cite{KIUMARSI20141167,vamvoudakis2021handbook}, and policy gradient (PG) \cite{gravell2019learning,fazel2018global,abbasi2019model,wang2023policy,cohen2018online,bu2019lqr}, as branches of reinforcement learning (RL), are typically data-hungry for stochastic systems. 
This is because they need many state-input realizations from the system to approximate the gradients in PG methods and the value functions in VL methods to cancel out or reduce the effect of noise. 

To avoid data hungriness, there has been a surge of interest in presenting one-shot and non-iterative direct data-based controllers for the LQR problem~\cite{DEPERSIS2021109548,de2019formulas,10061542,dorfler2022bridging}. This approach formalizes the data-driven control problem as a convex optimization problem, which can be solved at once. Although elegant, there are two drawbacks to these results. First, they cannot provide stability guarantees in the design time, and only a posterior stability analysis is provided. That is, a bound on the disturbance is obtained for which robust stability is guaranteed. However, this bound depends on the designed controller; thus, the bound and robust stability can only be verified after the design. Although a regularization term is added to the cost function to improve robustness, design time guarantees cannot be provided. Second, this regularization affects the trade-off between robustness and performance, and there is no systematic approach to tune it depending on the data quality.

{The motivation of this paper is to present one-shot data-efficient learning algorithms with design-time stability and performance guarantees to solve the direct data-driven discounted infinite-horizon LQR problem. The presented approach provides a new computationally efficient convex optimization formalization for the LQR problem under which the stability of the learned solution is guaranteed.} More specifically, the stochastic Lyapunov theory and Bellman inequality are leveraged to formalize a semidefinite programming (SDP) formulation for the data-driven LQR problem. 
{SDP is a convex program with efficient numerical solvers that allows the solution of large-scale problems that are formalized as linear matrix inequalities (LMIs).
To this end, an LMI-based solution to the LQR using a single trajectory of data is first formalized in which data-based closed-loop system dynamics appear. A control-oriented approach is then presented that characterizes the closed-loop system dynamics using data and a decision variable from which the control is obtained. That is, the decision variable, and thus the controller and closed-loop system, are learned to optimize the LQR cost by solving data-based LMIs. This closed-loop characterization is noise-dependent. In sharp contrast to \cite{DEPERSIS2021109548}, the presented formalization accounts for the noisy characterization of the closed-loop system through the Bellman inequality. As a consequence, a robustness term depending on the noise covariance naturally appears in this formulation, obviating the need to make a trade-off between performance and robustness manually. Therefore, robustness guarantees are provided in the design time by construction without adding any explicit regularization and, thus, without any need to hand-tune the trade-off between robustness and performance. The presented data-driven approach is a model-free learning method in which the optimal control policy is learned directly from the data by characterizing a data-based closed-loop system
without assuming or learning any explicit open-loop model of system dynamics.} 

{To guarantee the stability, it is shown that the learned direct data-driven closed-loop system in this model-free learning approach suffers from both multiplicative and additive noises. This is because the closed-loop system belongs to a distribution, and no point-based estimation is possible in the presence of additive noise.}
Robust direct data-driven LQR algorithm is designed to account for multiplicative and additive noises in the learned closed-loop systems. 
{
Due to the presence of noise and the stochasticity of the learned closed-loop system, from which a solution is found,  the performance of the learned policy may not achieve the
true optimal performance due to limitations in the learning algorithm or the data available. The difference between the
performance of the learned policy and the true optimal performance is called the suboptimality gap.
}
The analysis of the suboptimality gap reveals that if the signal-to-noise ratio (SNR) is low, the norm of the portion of the optimal control gain that belongs to the null space of the collected state data must be small. Also, it reveals that a small discount factor decreases the optimality gap but reduces the robustness to stability. Furthermore, it shows that the norm of the collected input signals used for excitation of the system has a direct impact on the performance gap. Simulation results on an active car suspension problem show the superiority of our approach in terms of robustness and performance gap compared to those 
{
of~\cite{DEPERSIS2021109548,treven2021learning,10061542}.}

This paper is organized as follows. Section II presents the model-based LQR problem formulation. Section III provides the data-driven LQR formalism. Section IV presents the Mean Square Stable (MSS) direct data-driven LQR approach. Section V analyzes the suboptimality gap between the model-based and the direct data-driven LQR costs. In Section VI, a simulation example is conducted to show the superiority of the proposed approach. Finally, section VII concludes the paper.  

\noindent \textbf{Notations and Definitions.} Throughout the paper, $\mathbb{R}$ denotes the real numbers. $Q (\preceq,\succeq)  0$ denotes that  $Q$ is (negative, positive) semi deﬁnite. $I$ denotes the identity matrix of the appropriate dimension. For a matrix $A$, $\text{Tr}(A)$ is its trace, $\rho(A)$ is its spectral radius, 
{
$\lambda_{min}(A)$ is its minimum eigenvalue, $\lambda_{max}(A)$ is its maximum eigenvalue,} and $A^\dag$ denotes its right inverse. All random variables are assumed to be defined on a probability space $(\Omega,\mathcal{F},\mathbb{P})$, with $\Omega$ as the sample space, $\mathcal{F}$ as its associated $\sigma$-algebra and $\mathbb{P}$ as the probability measure. For a random variable $w: \Omega \longrightarrow \mathbb{R}^n$ defined in the probability space $(\Omega,\mathcal{F},\mathbb{P})$, with some abuse of notation, the statement $w \in \mathbb{R}^n$ is used to state a random vector with dimension $n$. For a random vector $w$, $\mathbb{E}[w]$ indicates its mathematical expectation.
{$\mathcal{N}(0, W)$ represents the probability density function of a Gaussian distribution with mean $0$ and covariance matrix $W.$ $\mathbb{N}$ is the set of natural numbers.}\\
\begin{definition}{\cite{LAI2023110685,1103840}}
For a given control policy $u_k=h(x_k)$, the system (\ref{eq:syst}) is called MSS if there exists a positive definite $\Sigma_{\infty} \in \mathbb{R}^{n \times n}$ such that $\lim_{k \xrightarrow{}\infty}{\norm{\mathbb{E}(x_k x_k^T)-\Sigma_{\infty}}}=0$ and $\lim_{k \xrightarrow{}\infty}{\norm{\mathbb{E}(x_k)}=0}.$ In this case, the control policy is called admissible. 
\end{definition}
\section{Model-Based LQR Formulation}
This section is dedicated to model-based LQR formulation for linear time-invariant (LTI) stochastic discrete-time systems. We present two approaches to solve this problem. The first one is based on the dynamic programming principle that provides an analytical solution to the LQR problem~\cite{wang2015approximate}, \cite{LI2022110253}, and~\cite{10251494}. The second approach utilizes the stochastic Lyapunov stabilization theory to derive an optimization formulation for the LQR problem.  
\subsection{Problem Statement} \vspace{-4pt}
\noindent Consider an LTI stochastic discrete-time system of the form 
\begin{align} 
x_{k+1}=A \,x_k+B \, u_k+w_k,
\label{eq:syst}
\end{align} 
where $k \in \mathbb{N}$, $x_k \in \mathbb{R}^n$ is the system's state, and $u_k \in \mathbb{R}^m$ is the control input. Moreover, $A$ and $B$ are unknown transition and input matrices of appropriate dimensions, respectively. Furthermore, $w_k \in \mathbb{R}^n$ represents the system noise.
\begin{assumption}
The noise of the system $\omega_k$ is governed by the Gaussian distribution $\mathcal{N}(0, W)$ where $W \succ 0 \in \mathbb{R}^{n\times n}$ is its diagonal covariance matrix.   
\end{assumption}
\begin{assumption}
The pair $(A, B)$ is unknown but stabilizable.
\end{assumption}
\begin{assumption}
    The initial state $x_0 \in \mathbb{R}^n$ is assumed to be known. 
\end{assumption}
Throughout this paper, a linear control policy in the following form
\begin{align} 
u_{k}=K \,x_k,
\label{eq:control}
\end{align} 
 is used where $K \in \mathcal{K}$ is the control gain, and the set of stabilizing control gains is defined as $\mathcal{K} = \{K \in \mathbb{R}^{m \times n}: \rho(A+BK) < 1 \}.$

Consider the following cost function
\begin{align} 
J(K, x_0)=\sum_{k=0}^{\infty}{\gamma^k \mathbb{E}\left[x_k^T Q x_k + u_k^T R u_k\right]}
\label{eq:cost}
\end{align}
where $Q \in \mathbb{R}^{n \times n} \succ 0, R \in \mathbb{R}^{m \times m} \succ 0.$ Also, $\gamma \in (0,1)$ is the discount factor.\\
{
\begin{remark}
    The incorporation of the discount factor \(\gamma\) into the cost function \eqref{eq:cost} serves to mathematically limit the impact of process noise on the overall system cost. In the context of the infinite-horizon stochastic LQR problem, the discount factor is crucial for ensuring the finiteness of the cost. Specifically, setting \(\gamma = 1\) would result in an unbounded cost due to the noise terms, as given by \(\mathbb{E}\bigl[ \omega_k^\top (Q + K^\top R K) \omega_k \bigr] = \text{Tr}((Q + K^\top R K)W)\). Consequently,
    \begin{align*}
        \sum_{k=0}^{\infty} \text{Tr}((Q + K^\top R K)W) \rightarrow \infty.
    \end{align*}
    However, by employing a discount factor \(0 < \gamma < 1\), we confine this component of the cost function as 
    \begin{align*}
        \sum_{k=0}^{\infty} \gamma^k \text{Tr}((&Q + K^\top R K)W) = \nonumber \\
        &\frac{1}{1-\gamma} \text{Tr}((Q + K^\top R K)W) < \infty.
    \end{align*}
    Moreover, it is well-known that the discount factor influences the stability of the LQR problem. The stability analysis for the discounted LQR problem can be found in~\cite{9063479,LAI2023110685,LI2022110253}. In this paper, we assume a \(\gamma\) close to 1 that satisfies the stability conditions for the LQR solution, which will be provided later. 
\end{remark}
}
{
We utilize the following lemma from~\cite{LI2022110253} to establish a lower bound for the discount factor \(\gamma\) so that the system remains stable.
\begin{lemma} \label{lemma:gamma-stability}
    Let \( A \in \mathbb{R}^{n \times n} \). Then, for \(\gamma > 1 - \frac{\lambda_{\min}(M)}{\lambda_{\max}(A^\top P A)}\), the spectral radius \(\rho(A) < 1\) if and only if, for each given positive definite matrix \(M \in \mathbb{R}^{n \times n}\), there exists a unique positive definite matrix \(P \in \mathbb{R}^{n \times n}\) such that \(\gamma A^\top P A - P = -M\).\\
    \textbf{Proof}.Please refer to \cite{LI2022110253}.\qed
\end{lemma}
}
{
\begin{remark}
    Another approach to limit the infinite-horizon stochastic LQR cost is to consider the averaged cost as 
    \begin{align}
        J_{avg}(K) = \lim_{N \rightarrow \infty} {\frac{1}{N}J(K, x_0, N)}=Tr(PW),
    \end{align}
    where $J(K, x_0, N)$ is the horizon-dependent cost 
    \begin{align}
        J(K, x_0, N) = \sum_{k=0}^{N}{ \mathbb{E}\left[x_k^T Q x_k + u_k^T R u_k\right]}. \label{eq:finitecost}
    \end{align}
    It is clear that as $N$ goes to infinity, so does the cost \eqref{eq:finitecost}.
    It should be pointed out that $J_{avg}$ does not depend on $x_0$ and is only affected by noise covariance $W.$ The optimal cost of stochastic discounted infinite-horizon LQR \eqref{eq:Vstar} is highly dependent on the term $Tr(PW)$. As a result, the trace term is the dominant factor in both the discounted and averaged versions of the cost \eqref{eq:finitecost} in the infinite-horizon setting. This is true for deterministic and stochastic initial conditions. In Lemma \ref{model-based-lqr-lmi}, we show that the upper bound of the discounted infinite-horizon cost \eqref{eq:cost} also depends on the trace term $Tr(PW)$.
\end{remark}
}
\begin{assumption}
For the system (1) and the cost function (3), the pair $(A, \sqrt{Q})$ is detectable. 
\end{assumption}

\begin{problem}
    The LQR problem is to design the gain matrix $K$ in (\ref{eq:control}) for the system (\ref{eq:syst}) under Assumptions 1-4 to minimize the cost function (\ref{eq:cost}). 
\end{problem}

Based on the cost function defined in \eqref{eq:cost}, the cost-to-go function for the control policy (\ref{eq:control}) can be defined as 
\begin{align}
     V_K(x_k) &= \sum_{t=k}^{\infty} \gamma^{t-k}  
     \mathbb{E}\bigl[ x_t^T Q x_t + u_t^T R u_t 
       | x_k
       \bigr]. \label{eq:Vk}
\end{align}

\begin{lemma} \label{optimal-value-function}
Consider the LQR problem (\ref{eq:cost}). The optimal value function is given by
\begin{align}
    V^*(x_k) = \min_{K}{V_K(x_k)} = x_k^T  P^* x_k + c^*, \label{eq:Vstar}
\end{align}
where, 
\begin{align}
    &P^* = Q+K^{*^T} R K^* + \gamma(A+BK^*)^T P^* (A+BK^*), \label{eq:Pstar}\\
    &c^* = \frac{\gamma}{1-\gamma}Tr(P^* W), \label{eq:cstar}
\end{align}
and the optimal control gain $K^*$ has the following form
\begin{align}
    K^* = -\gamma (R+\gamma B^TP^* B)^{-1}B^TP^* A. \label{eq:Kstar}
\end{align}
\textbf{Proof}.See \cite{LI2022110253} and~\cite{wang2015approximate}.\qed
\end{lemma}
Substituting (\ref{eq:Kstar}) into (\ref{eq:Pstar}) results in the well-known Algebraic Ricatti equation (ARE), which can be solved to find $P^*$ and subsequently $K^*.$ Since ARE is quadratic in $P^*$, and thus hard to solve directly, iterative approaches such as policy iteration \cite{bertsekas2011approximate,hu2023toward,NEURIPS2019_aaebdb8b} have been used to solve it by iterating on Lyapunov equations to evaluate policies (policy evaluation) and then find an improved policy based on this evaluation (policy improvement). Data-driven implementations of these iterative algorithms have also been presented \cite{9691800}. However, a significant amount of interactive data must be collected (i.e., a batch of data for every policy under evaluation at every iteration). Although off-policy evaluation can reduce the sample complexity (i.e., the number of samples that must be collected to achieve a specific performance), it leads to a huge variance in performance estimation. In addition, due to noise, many realizations of the system must be collected (typically starting from the same initial conditions) \cite{9691800} to learn the control policy, which makes them data-hungry. 

To provide a low sample complexity data-driven algorithm for solving the LQR \textbf{Problem 1} using only one batch of data, a different approach is provided to find $P^*$ and $K^*$. That is, the LQR \textbf{Problem 1} is cast as SDPs or LMIs using Lyapunov theory (or Bellman inequality). Although SDP formulations of the LQR problem have been considered in \cite{DEPERSIS2021109548} based on the controllability Gramian of the closed-loop system, to provide stability robustness and improve performance gaps, we provide a new SDP formulation in the following subsection which makes it more intuitive to leverage the Bellman inequality to account for the robustness in the design time (rather than a posteriori analysis of stability) by representing the closed-loop dynamic as a system with multiplicative and additive noises. 
\subsection{LMI Formulation of Model-Based LQR Problem} 
This subsection presents an LMI formulation of the LQR problem, which will be leveraged to design one-shot data-driven solutions for the LQR \textbf{Problem 1}. The approach is based on applying the Lyapunov theory in consecutive steps. Lemma 3 formalizes the LMI approach to the LQR \textbf{Problem 1}. \vspace{6pt}
{
\begin{lemma} \label{model-based-lqr-lmi}
    Consider the system (\ref{eq:syst}) with the control policy (\ref{eq:control}). Let Assumptions 1-3 hold and 
 $x_0$ be a given initial condition. Then, the optimal state feedback matrix $K$ that minimizes the cost function (\ref{eq:cost}) of the LQR \textbf{Problem 1} is given by
    \begin{align}
        K = M Y^{-1}, 
    \end{align}
    where $P=Y^{-1} \succ 0$ and $M \in \mathbb{R}^{m \times n}$ are the solutions (if they exist) to the following problem
\begin{subequations}
 \label{eq:optlmiModel}
\begin{align}
    \max_{M, Y} \quad & {Tr(W^{-1}Y)}\\
\textrm{s.t.} \quad &  \begin{bmatrix}
        -Y & Y & M^T & (AY+BM)^T  \\
        * & -Q^{-1} & 0 & 0 \\
        * & * & -R^{-1} & 0 \\
        * & * & * & -\frac{Y}{\gamma} 
        \end{bmatrix} \preceq 0, \label{eq:modelLMI} \\
        \quad & Y \succeq 0.
\end{align}
\end{subequations}
provided that 
\begin{align}
    \gamma > 1- \frac{\lambda_{min}(Q + K^\top R K)}{\lambda_{max}((A+BK)^\top P(A+BK))}. \label{eq:model-gamma-lower}
\end{align}
Furthermore, the closed-loop system under the optimal control policy is MSS, and the resulting policy is admissible. \\
\textbf{Proof}. The proof is performed by finding an upper bound of the cost function (\ref{eq:cost}) and minimizing this upper bound, which gives the optimal solution. \newline
Let the modified system $x_{k+1}=\sqrt{\gamma}A x_k+\sqrt{\gamma}B u_k+\sqrt{\gamma}w_k$  be defined  by replacing $A$ and $B$ with $\sqrt{\gamma}A$ and $\sqrt{\gamma}B$ in  the system (\ref{eq:syst}), respectively. According to the Lyapunov stability theorem for stochastic systems~\cite{HADDAD2022110393}, suppose there exists a Lyapunov function $\bar{V}_K(x_k)=x_k^T P x_k+c,$ for the modified system, with $P \succeq 0 $, and $c \geq 0$ being a constant that is defined later. Then, the following Lyapunov inequalities hold
\begin{align}
    \gamma \mathbb{E}\bigl[x_{1}^T P x_1 + c \bigr]-x_0^T P x_0 - c  & \leq -(x^T_0 Q x_0 + u^T_0 R u_0) \nonumber \\ 
    \gamma \mathbb{E}\bigl[x_{2}^T P x_2 + c\bigr]-x_1^T P x_1 - c & \leq -(x^T_1 Q x_1 + u^T_1 R u_1) \nonumber \\
    &\vdots \nonumber \\
    \gamma \mathbb{E}\bigl[x_{k+1}^T P x_{k+1} + c\bigr]-x_k^T P x_k  - c& \leq -(x^T_k Q x_k + u^T_k R u_k) \nonumber \\
    \vdots  \label{eq:Vlmi}
\end{align}
where $u_k = K x_k$ for $\forall k \in \mathbb{N}$. By multiplying both sides of the inequalities by $\gamma^k$, summing up all the inequalities, and taking the expectation with respect to the noise probability distribution, the following inequality holds 
\begin{dmath}
    -x_0^T P x_0 - c - \sum_{k=1}^{\infty}{\gamma^k Tr(P(\bar{\omega}_k \bar{\omega}_k^T - W))} \leq -\sum_{k=0}^{\infty}{\gamma^k \mathbb{E}\bigl[ x^T_k Q x_k + u^T_k R u_k \bigr]}. \label{eq:JV}
\end{dmath}
if $\lim_{k \xrightarrow{} \infty} {\gamma^{k+1} \mathbb{E} \bigl[ x^T_{k+1}Px_{k+1}} \bigr] = 0$ is satisfied, where $\bar{\omega}_k$ is a realization of $\omega_k.$  Using tail bounds for the chi-squared distribution, there exists a $0 < \beta < \infty$ for which it holds with some probability that $\bar{\omega}_k \bar{\omega}_k^T \preceq \beta W$ and 
\begin{align}
   Tr(P \bar{\omega}_k \bar{\omega}_k^T) - Tr(PW) \leq (\beta -1)Tr(PW). 
\end{align} \\
Based on Lemma \ref{lemma:gamma-stability}, the condition $$\lim_{k \xrightarrow{} \infty} {\gamma^{k+1} \mathbb{E} \bigl[ x^T_{k+1}Px_{k+1}} \bigr] = 0$$  leads to $\mathbb{E}\bigl[x_\infty\bigr]=0$ if $\gamma$ satisfies the condition \eqref{eq:model-gamma-lower}. That is, the system is MSS and based on (\ref{eq:cost}), (\ref{eq:Vk}), (\ref{eq:JV}), and considering the worst-case for the realizations of $\bar{\omega}_k$, one has the following cost upper bound 
\begin{align}
    J(K, x_0) \leq \bar{V}_K(x_0) + \frac{(\beta -1)\gamma}{1-\gamma}Tr(PW), \label{eq:JV0}
\end{align}
for the control gain $K$ and the original system \eqref{eq:syst}.
 From any of the Lyapunov inequalities in \eqref{eq:Vlmi}, one can extract the following condition for $c$
\begin{align}
    c = \frac{\gamma}{1-\gamma}Tr(PW), \label{eq:c_lyap}
\end{align}
which results in the following upper bound for the LQR cost 
\begin{align}
J(K, x_0) \leq x_0^\top P x_0 + \frac{\beta\gamma}{1-\gamma}Tr(PW).
\end{align} \\
When $\beta=1$ (or the expected upper bound, that is, $\mathbb{E}[\bar{\omega}_k \bar{\omega}_k^\top]=W$) then $\bar{V}_K(x_0)$ is a probable upper bound for the model-based LQR cost (\ref{eq:cost}). Minimizing this bound leads to the finding of the optimal controller that optimizes the cost function (\ref{eq:cost}). 
This probable upper bound is dominated by the second term $\frac{\gamma}{1-\gamma}Tr(PW)$ due to the weighting factor $\frac{\gamma}{1-\gamma}$, which captures the effect of the noise on the long-term cost of the system.  Since the first term of this upper bound cost also depends on $P$, which is weighted by the initial conditions, the overall upper bound term $x_0^T P x_0 +\frac{\gamma}{1-\gamma}Tr(PW)$ can be safely replaced with $Tr(PW)$. \\
The upper bound must be optimized for admissible policies. The condition for the lower bound of $\gamma$ \eqref{eq:model-gamma-lower} that ensures admissibility can be obtained using Lemma \ref{lemma:gamma-stability}. An admissible policy $K$ and a value matrix $P$ must also satisfy every inequality in (\ref{eq:Vlmi}) which is equivalent to 
\begin{align}
    \gamma(A+BK)^T P(A+BK) - P + Q + K^T R K \preceq 0.
\end{align}
This condition can be translated into the LMI (\ref{eq:modelLMI}) using the Schur complement twice. Finally, instead of directly minimizing $Tr(PW)$, the inequality 
\begin{align}
    \frac{1}{Tr(W^{-1}Y)} < \frac{Tr(PW)}{n^2}, \label{eq:PYineq}
\end{align}
is leveraged, leading to maximize the term $Tr(W^{-1}Y).$ In other words, the lower bound of $V_K(x_0)$ is minimized by optimizing over admissible $K$ and $P.$ This completes the proof. 
\qed
\end{lemma}
\begin{remark}
    Since we minimize the upper bound of the cost, it provides robustness to noise. The emphasis here is on the expected performance over time, considering the stochastic nature of the system. The initial condition does not affect the overall strategy for finding the optimal control solution. It should be pointed out that minimizing the upper bound of the cost function 
$J(K, x_0)$ leads to finding the optimal controller~\cite{MARTINELLI2022110052}. 
\end{remark}
\begin{remark}
    In the infinite-horizon stochastic LQR problem, the discount factor \(\gamma\) is not employed as a tuning parameter and rather is used to ensure the mathematical soundness of the problem. To ensure that the lower bound for \(\gamma\) given in \eqref{eq:model-gamma-lower} is satisfied, we utilize a large value close to 1 for \(\gamma\).
\end{remark}
\begin{remark}
     The term $Tr(PW)$ optimized in Lemma 3 also appears within the data-driven LMI condition, as shown later. For the case where the initial condition \(x_0\) is a random variable independent of \(\omega_k\) and distributed as \(x_0 \sim \mathcal{N}(\bar{x}_0, \Sigma)\), the upper bound of the cost is given by
    \begin{align}
        J(K, x_0) \leq \bar{x}_0^\top P \bar{x}_0 + \text{Tr}(P\Sigma) + \frac{\gamma}{1 - \gamma} \text{Tr}(PW).
    \end{align}
    When the uncertainty of the initial condition \(\Sigma\) is comparable to \(\frac{\gamma}{1 - \gamma} W\), it becomes necessary to minimize the entire upper bound to optimize performance. 
\end{remark}
}
\section{Data-based LQR Formulation}
As LMIs provide an elegant one-shot approach to designing the LQR gain, it is beneficial to formulate data-driven counterparts of the LMI formulation presented in Lemma 2. 
Suppose that we have collected $N$ sequences of data as
\begin{subequations}
\label{eq:collected_data}
\begin{align} 
    U_0 &= \begin{bmatrix}
        u_0 & u_1 & \cdots & u_{N-1}
    \end{bmatrix}, \\
    X_0 &= \begin{bmatrix}
        x_0 & x_1 & \cdots & x_{N-1}
    \end{bmatrix}, \\
    X_1 &= \begin{bmatrix}
        x_1 & x_2 & \cdots & x_{N}
    \end{bmatrix},
\end{align}
\end{subequations}
where $U_0 \in \mathbb{R}^{m \times N}$, $X_0$ and $X_1 \in \mathbb{R}^{n \times N}$. The corresponding noise sequences $\Omega_0 \in \mathbb{R}^{n \times N}$ are 
\begin{align}
    \Omega_0 = \begin{bmatrix}
        \omega_0 & \omega_1 & \cdots & \omega_{N-1}
    \end{bmatrix},
\end{align}
in which we have no access to {and we treat its columns as Gaussian random variables with mean $0$ and covariance $W$.
Based on the system dynamics (1), the collected data satisfies 
\begin{align}
    X_1 = A X_0 + B U_0 + \Omega_0. \label{eq:data-integrity}
\end{align}
}
\\
\begin{assumption}
The collected data matrix $D_0 = \begin{bmatrix}
    U_0^T & X_0^T
\end{bmatrix}^T$ is sufficiently rich, i.e., it has full row rank. That is,
\begin{align}
    rank(D_0) = m+n. \label{eq:rank}
\end{align}
\end{assumption}
The collected sequences show that the pair of unknown system matrices $(B, A)$ belongs to the uncertain set defined by 
\begin{align}
    \mathcal{S} = \{(B, A) \,\, | \,\,   X_1 = \begin{bmatrix}
        B & A
    \end{bmatrix} D_0 + \Omega_0,  \nonumber  \\
    \, \omega_i \sim \mathcal{N}(0, W), \, i=1,\cdots,N-1 \}.
\end{align}
Based on the presented data-based formulation, in the upcoming subsections, we will discuss indirect and direct LMI formulations of data-driven LQR problems in the literature.
\subsection{Indirect data-driven (CE approach)}
In general, indirect data-driven control is obtained by a model-based control design after a system identification procedure that provides a system model. In the stochastic case, the identified model is one of the models in the set $\mathcal{S},$ as it depends on the collected data (\ref{eq:collected_data}).
Consider the following lemma, which is the CE counterpart of Lemma 2. 
\begin{lemma} \label{indirect-data-based-lqr-lmi}
    Consider the system (\ref{eq:syst}) and let the data collected from this system be given as (\ref{eq:collected_data}). Let Assumptions 1-4 hold and 
 $x_0$ be a given initial condition. Then, the CE state feedback matrix $K$ that solves the LQR \textbf{Problem 1} is given by $K = M Y^{-1}, $
    where $P=Y^{-1} \succ 0$ and $M \in \mathbb{R}^{m \times n}$ are the solutions (if they exist) to the following bi-level optimization problem\\
\begin{subequations}
\label{eq:celmi}
\begin{align} 
    \max_{M, Y} \quad & {Tr(W^{-1}Y)}\\
\textrm{s.t.} \quad &  \begin{bmatrix}
        -Y & Y & M^T & (\hat{A}Y+\hat{B}M)^T  \\
        * & -Q^{-1} & 0 & 0 \\
        * & * & -R^{-1} & 0 \\
        * & * & * & -\frac{Y}{\gamma} 
        \end{bmatrix} \preceq 0, \\
        \quad & Y \succeq 0, \\
        \quad & \begin{bmatrix}
            \hat{B} &\hat{A}
        \end{bmatrix} = \argmin_{B, A} \lVert X_1-\begin{bmatrix}
            B & A
        \end{bmatrix}D_0 \rVert_F, \label{eq:BAh}
\end{align}
\end{subequations}
where $\lVert . \rVert_F$ represents the Frobenius norm.
{ $\hat{B}$ and $\hat{A}$ are the estimates of $B$ and $A$, respectively.}
\\
\textbf{Proof. } The equation (\ref{eq:BAh}) solves the least-squares (LS) problem as~\cite{10061542}
\begin{align*}
    \begin{bmatrix}
            \hat{B} &\hat{A}
        \end{bmatrix} = \argmin_{B, A} \lVert X_1-\begin{bmatrix}
            B & A
        \end{bmatrix}D_0 \rVert_F = X_1 D_0^{\dag},
\end{align*}
which has a solution under Assumption 4.  If we substitute $\hat{A}$ and $\hat{B}$ in the problem (\ref{eq:optlmiModel}), then the indirect data-driven LQR (\ref{eq:celmi}) will result. \qed
\end{lemma}

It is known that if the data is noise-free and the condition (\ref{eq:rank}) holds, then Problem (\ref{eq:celmi}) is feasible and returns the optimal control gain $K=MP.$ In our case, the data is noisy, so this problem does not produce the optimal control, and it also needs to be robustified as the identified matrices $\hat{A}$ and $\hat{B}$ may not be the same as actual system matrices $A$ and $B$ respectively~\cite{9992770}. A common approach is to estimate an ellipsoidal uncertainty set for the system matrices using the collected data and then incorporate this knowledge into the controller design using S-procedure~\cite{treven2021learning,vien2021differentiable}.

{
\begin{remark}
    In the absence of prior knowledge of the system matrices \(A\) and \(B\), the noise covariance matrix \(W\) can be estimated from the collected data using an empirical approach. The procedure involves the following steps:
    \begin{enumerate}
        \item \textbf{Estimate \(A\) and \(B\)}: Solve the LS problem \eqref{eq:BAh} to estimate the matrices \(A\) and \(B\). This can be achieved using standard linear regression techniques.
        \item \textbf{Compute Residuals}: Using the estimated matrices \(\hat{A}\) and \(\hat{B}\), compute the residuals
        \[
        \hat{w}_k = x_{k+1} - \hat{A} x_k - \hat{B} u_k, \quad  \ k = 0, 1, \ldots, N-1.
        \]
        \item \textbf{Estimate Noise Covariance}: The noise covariance matrix \(W\) is then estimated ($\hat{W}$) using the sample covariance of the residuals
        \[
        \hat{W} = \frac{1}{N} \sum_{k=0}^{N-1} \hat{w}_k \hat{w}_k^\top.
        \]
    \end{enumerate}
    This method leverages the data to estimate the system matrices and, subsequently, the noise covariance, ensuring that the control process is grounded in observed data.
\end{remark}
}
\subsection{Direct data-driven}
Direct data-driven control aims to eliminate the system identification step (at least explicitly) and design the controller directly using the collected data (\ref{eq:collected_data}). One way is to parameterize the closed-loop system matrices using the collected data and then design a controller for it. This involves learning the closed-loop dynamics. By employing the parameterization of $K$ as proposed in~\cite{DEPERSIS2021109548}, for any $K$ there is a matrix $G \in \mathbb{R}^{N \times n}$ such that
\begin{align}
    \begin{bmatrix}
        K \\ I
    \end{bmatrix} = D_0 G. \label{eq:KD0G}
\end{align}
{
Multiplying \eqref{eq:data-integrity} by $G$ from right yields
\begin{align}
    X_1 G = A X_0 G + B U_0 G + \Omega_0 G. \label{eq:x1g}
\end{align}
By combining \eqref{eq:x1g} with \eqref{eq:KD0G}, one has
}
\begin{align}
    A+BK &= (X_1 - \Omega_0) G, \label{eq:ABK}\\
    X_0 G &= I, \label{eq:X0GI}
\end{align}
{
which provides a data-based representation of the closed-loop transition matrix.
}
In general, some part of the system matrices defined by $\mathcal{S}$ belongs to the set of closed-loop matrices 
\begin{align}
    \mathcal{S_K} = \{A+BK \, | \, (B, A) \in \mathcal{S} \}.
\end{align}
that are consistent with the collected data~\cite{9147320}.
The following lemma describes this approach.
\begin{lemma}
    Consider the system (\ref{eq:syst}) and let the data collected from this system be given as (\ref{eq:collected_data}) and let Assumptions 1-4 hold. Then, the direct CE LQR control gain $K = U_0 G$ with $G=FY^{-1}$ and $Y=P^{-1}$  solves the following direct data-based optimization problem
    \begin{subequations}
\label{eq:directCelmi}
\begin{align} 
    \max_{F, Y} \quad & {Tr(W^{-1}Y)}\\
\textrm{s.t.} \quad &  \begin{bmatrix}
        -Y & Y & (U_0 F)^T & (X_1 F)^T\\
        * & -Q^{-1} & 0 & 0\\
        * & * & -R^{-1} & 0 \\
        * & * & * & -\frac{Y}{\gamma}
        \end{bmatrix} \preceq 0, \\
        \quad & X_0 F = Y, \\
        \quad & Y \succeq 0.
\end{align}
\end{subequations}
\textbf{Proof.} By substituting the parameterization in (\ref{eq:ABK}) and (\ref{eq:X0GI}) into the problem (\ref{eq:optlmiModel}), the following direct formulation of the LQR problem (no identification) is achieved
\begin{subequations}
\label{eq:directlmi}
\begin{align} 
    \max_{F, Y} \quad & {Tr(W^{-1}Y)}\\
\textrm{s.t.} \quad &  \begin{bmatrix}
        -Y & Y & (U_0 F)^T & ((X_1-\Omega_0) F)^T\\
        * & -Q^{-1} & 0 & 0\\
        * & * & -R^{-1} & 0 \\
        * & * & * & -\frac{Y}{\gamma}
        \end{bmatrix} \preceq 0, \\
        \quad & X_0 F = Y, \\
        \quad & Y \succeq 0.
\end{align}
\end{subequations}
\noindent The existence of a solution is guaranteed under Assumptions 1-4. Since the noise data $\Omega_0$ is not measurable, the direct CE~\cite{DEPERSIS2021109548} optimization problem (\ref{eq:directCelmi}) is obtained by disregarding the noise data $\Omega_0$ from \eqref{eq:directlmi}, which results in \eqref{eq:directCelmi}.\qed
\end{lemma}
The problem (\ref{eq:directCelmi}) suffers the robustness feature since we disregarded the noise data. One way to alleviate this issue is to force a specific solution from the null space of the data matrix $D_0$ by adding the following orthogonality constraint 
\begin{align}
    (I-D_0^{\dag} D_0)G = 0, \label{eq:orthogonality}
\end{align}
as a regularizer to Problem (\ref{eq:directCelmi}). It has been shown in~\cite{DEPERSIS2021109548,9992770} that the indirect data-driven control Problem (\ref{eq:celmi}) is equivalent to the direct data-driven control Problem (\ref{eq:directCelmi}) with the regularization term (\ref{eq:orthogonality}) with large enough weight. Another approach to robustify the direct CE LQR is to regularize the cost function by adding stability-enhancing criteria, see Remark 4.1 in~\cite{10061542}.
\vspace{1pt}\\
\begin{remark}
There are some challenges associated with the regularization approaches mentioned above. It is difficult to tune the regularization parameter to balance between optimality and robustness. Also, they are ad-hoc methods and lack stability analysis; only posterior robustness analysis is provided.
\end{remark}
The next section provides a remedy for both problems mentioned in Remark 7. So, unlike ~\cite{10061542} and \cite{DEPERSIS2021109548}, there is no extra tunable parameter to promote robustness since it is done implicitly by construction. Also, we do not neglect the effect of immeasurable noises $\Omega_0$ on deriving the new formulation for direct data-driven LQR.

\section{MSS Direct Data-Driven LQR}
When learning the dynamics for the sake of control, as equations (\ref{eq:ABK}) and (\ref{eq:X0GI}) imply, one deals with an uncertain closed-loop dynamic which is affected by the noise matrix $\Omega_0$. In other words, a stochastic closed-loop dynamic (i.e. a family of dynamics) is obtained and must be stabilized.  Lemma \ref{lemma:multiplicative} formalizes this fact. 
\begin{lemma} \label{lemma:multiplicative}
The following direct data-driven closed-loop system formulation 
\begin{align}
    x_{k+1} = (X_1 G- \Omega_0G) x_k + \omega_k. \label{eq:closedLoopDirect}
\end{align}  
represents a system with multiplicative and additive noises as
\begin{align}
    x_{k+1} = (A_0 + \sum_{i=1}^{nN}{\vartheta_i}A_i)x_k + \omega_k, \label{eq:multiSys}
\end{align}
where $A_0=X_1G$ and $\sum_{i=1}^{nN}{\vartheta_i}A_i=-\Omega_0 G$. In this formulation, $\vartheta_i$ is the $i-$th element of 
\begin{align}
    {\vartheta} = \begin{bmatrix}
        \omega_0^T & \omega_1^T & \cdots & \omega_{N-1}^T
    \end{bmatrix},
\end{align}
where $\omega_t^T = \begin{bmatrix}
        \omega_t^{(1)} & \cdots & \omega_t^{(j)} & \cdots & \omega_t^{(n)}
    \end{bmatrix}$ and as a result $\vartheta_i = \omega_t^{(j)}$, for example $\vartheta_1=\omega_0^{(1)}$ and $\vartheta_{n+1}=\omega_1^{(1)}.$
The matrix $A_i$ is an $n\times n$ matrix where all of its rows are zero vectors except the $j-$th row which is equal to the negative of the $(t+1)-$th row of $G$. \\
{
\textbf{Proof.} The proof is based on matrix multiplication and some algebraic calculations. First, consider the simple case of $n=2$ and $N=3$
\begin{align}
    \Omega_0 = \begin{bmatrix}
        \omega_0^{(1)} & \omega_1^{(1)} & \omega_2^{(1)} \\
        \omega_0^{(2)} & \omega_1^{(2)} & \omega_2^{(2)}
    \end{bmatrix}, G = \begin{bmatrix}
        g_{11} & g_{12} \\ g_{21} & g_{22} \\ g_{31} & g_{32}
    \end{bmatrix}, \nonumber
\end{align}
then
$
    {\vartheta} = \begin{bmatrix}
        \omega_0^{(1)} & \omega_0^{(2)} & \omega_1^{(1)} &
        \omega_1^{(2)} & \omega_2^{(1)} & \omega_2^{(2)}
    \end{bmatrix}, 
$
and 
\begin{align}
    A_1 = -\begin{bmatrix}
        g_{11} & g_{12} \\ 0 & 0
    \end{bmatrix}, A_2 = -\begin{bmatrix}
        0 & 0 \\ g_{11} & g_{12}
    \end{bmatrix}, \nonumber \\
    A_3 = -\begin{bmatrix}
        g_{21} & g_{22} \\ 0 & 0
    \end{bmatrix}, A_4 = -\begin{bmatrix}
        0 & 0 \\ g_{21} & g_{22}
    \end{bmatrix}, \nonumber \\
    A_5 = -\begin{bmatrix}
        g_{31} & g_{32}\\ 0 & 0
    \end{bmatrix}, A_6 = -\begin{bmatrix}
        0 & 0 \\ g_{31} & g_{32}
    \end{bmatrix}. \nonumber
\end{align}
The same procedure can be easily generalized to the case where $n>2$ and $N>3$ by finding the result of $\Omega_0 G$ multiplication using the following general forms
\begin{align}
    &\Omega_0 G = \begin{bmatrix}
        \omega_0^{(1)} & \omega_1^{(1)} & \cdots & \omega_{N-1}^{(1)} \\
        \omega_0^{(2)} & \omega_1^{(2)} & \cdots & \omega_{N-1}^{(2)} \\
        \vdots & \vdots & \ddots & \vdots \\
        \omega_0^{(n)} & \omega_1^{(n)} & \cdots & \omega_{N-1}^{(n)}
    \end{bmatrix}\begin{bmatrix}
        g_{11} & g_{12} & \cdots & g_{1n} \\
        g_{21} & g_{22} & \cdots & g_{2n} \\
        \vdots & \vdots & \ddots & \vdots \\
        g_{N1} & g_{N2} & \cdots & g_{Nn}
    \end{bmatrix} =\nonumber \\ 
    &\omega_0^{(1)}\begin{bmatrix}
        g_{11} & g_{12} & \cdots & g_{1n} \\
        0 & 0 & \cdots & 0 \\
        \vdots & \vdots & \ddots & \vdots \\
        0 & 0 & \cdots & 0
    \end{bmatrix}+\cdots+\omega_0^{(n)}\begin{bmatrix}
        0 & 0 & \cdots & 0 \\
        \vdots & \vdots & \ddots & \vdots \\
        0 & 0 & \cdots & 0\\
        g_{11} & g_{12} & \cdots & g_{1n} \\
    \end{bmatrix} \nonumber \\
   &+\cdots+\omega_{N-1}^{(n)}\begin{bmatrix}
    0 & 0 & \cdots & 0 \\
        \vdots & \vdots & \ddots & \vdots \\
        0 & 0 & \cdots & 0 \\
        g_{N1} & g_{N2} & \cdots & g_{Nn} 
    \end{bmatrix}=-\sum_{i=1}^{nN}{\vartheta_i}A_i.
\end{align}
\qed
}
\end{lemma}

Stability analysis of multiplicative noise systems has been extensively studied in ~\cite{GRAVELL20207392,coppens2022safe,LAI2023110685,1103840,pmlr-v120-coppens20a}.  
\begin{theorem}
    Consider the system (\ref{eq:syst}) and let the data collected from this system be given as (\ref{eq:collected_data}) and Assumptions 1-4 hold. Then, the robust LQR control gain $K = U_0 G$ with $G=FY^{-1}$ and $Y=P^{-1}$  solves the following data-based optimization problem
\begin{subequations}
\label{eq:directRobustlmi}
\begin{align}
 \max_{F, Y, \alpha} \quad & \alpha\\
\textrm{s.t.} \quad & \begin{bmatrix}
        -Y & Y & (U_0 F)^T & (X_1 F)^T & F^T \\
        * & -Q^{-1} & 0 & 0 & 0\\
        * & * & -R^{-1} & 0 & 0 \\
        * & * & * & -\frac{Y}{\gamma} & 0 \\
        * & * & * & * & -\frac{\alpha}{ \gamma}I
        \end{bmatrix} \preceq 0. \label{eq:lmiDD} \\
        \quad & X_0F=Y, \\
        \quad &Tr\left(W^{-1}Y\right)-\alpha n^2\ge 0, \label{eq:tracePW} \\
        \quad & \alpha > 0,
\end{align}
\end{subequations}
{
provided that 
\textcolor{blue}{
\begin{align}
    \gamma > 1- \frac{\lambda_{min}(Q + (U_0 G)^\top R (U_0 G))}{\lambda_{max}((X_1 G)^\top P (X_1 G) + Tr(PW) G^\top G)}. \label{eq:data-gamma-lower}
\end{align}}
}
Furthermore, the LQR policy $u_k = U_0 G x_k$ is admissible and the closed-loop system is MSS.\\
\textbf{Proof.} 
{
Based on the proof of Lemma 3, equations (\ref{eq:KD0G})-(\ref{eq:X0GI}), and the data-driven closed-loop system \eqref{eq:closedLoopDirect} or \eqref{eq:multiSys}, the expected upper bound of the direct data-driven version of the LQR problem is in the same form $x_0^\top P x_0 + \frac{\gamma}{1-\gamma}Tr(PW)$ as the model-based one. As a result, we can formulate the direct data-driven LQR problem as
\begin{subequations}
\label{eq:robustform}
\begin{align} 
    \min_{G, P} \quad & {Tr(PW)}\\
\textrm{s.t.} \quad &  x_k^\top P x_k + c \ge  x_k^\top Q x_k + (U_0 G x_k)^\top R (U_0 G x_k)  \nonumber \\ 
&+ \gamma \mathbb{E}\left[ x_{k+1}^\top P x_{k+1} + c\right],  \forall k \in \mathbb{N},\label{eq:ddbell} \\
\quad & X_0 G = I, \\
        \quad & P \succ 0. 
\end{align}
\end{subequations}
in which conditions (\ref{eq:ddbell}) are the data-based counterpart of conditions (\ref{eq:Vlmi}). The optimal cost of Problem (\ref{eq:robustform}) is an upper bound of the LQR cost (\ref{eq:cost}). Therefore, minimizing this upper bound provides a way to reach the LQR value function and optimal control gain. Using \eqref{eq:closedLoopDirect} or (\ref{eq:multiSys}), the direct data-driven Bellman inequalities (\ref{eq:ddbell}) can be written as 
\begin{dmath}
      x_k^\top (Q+(U_0 G)^\top R U_0 G)x_k + 
    \gamma \mathbb{E}\left[ \left((A_0 + \sum_{i=1}^{nN}{\vartheta_i}A_i)x_k + \omega_k\right)^\top P \left((A_0 + \sum_{i=1}^{nN}{\vartheta_i}A_i)x_k + \omega_k\right) + c\right] \leq x_k^\top P x_k + c , \label{eq:DDDD}
\end{dmath}
for $\forall k \in \mathbb{N}$. It is satisfied if $c$ is chosen as \eqref{eq:c_lyap} and
\begin{dmath}
  Q+(U_0 G)^\top R U_0 G + 
    \gamma \mathbb{E}\left[ \left(A_0 + \sum_{i=1}^{nN}{\vartheta_i}A_i\right)^\top P \left(A_0 + \sum_{i=1}^{nN}{\vartheta_i}A_i\right)\right] \leq P. \label{eq:DDDDnoxk}
\end{dmath}
Since we do not have access to $\Omega_0$, we treat its columns as Gaussian random variables with mean $0$ and covariance $W$. As a result, the expectation is taken with respect to all random variables i.e. $\omega_k$ and $\Omega_0.$
After taking the expectation over noise density $\vartheta$ and some algebraic simplifications on (\ref{eq:DDDDnoxk}) as
\begin{dmath}
\mathbb{E}\left[ \left( \sum_{i=1}^{nN}{\vartheta_i}A_i) \right)^T P \left( \sum_{i=1}^{nN}{\vartheta_i}A_i) \right) \right] = \mathbb{E}\left[ (\Omega_0 G)^\top P \Omega_0 G\right] = G^\top \mathbb{E}\left[ \Omega_0^\top P \Omega_0 \right]G = G^\top Tr(PW)I G = Tr(PW) G^T G,
\end{dmath}
}
the following matrix inequality results
{
\begin{align}
&P-\gamma (X_1 G)^\top P X_1 G - Q-(U_0 G)^\top R U_0 G \nonumber \\ 
& \quad -\gamma Tr(PW)G^\top G \succeq 0. \label{eq:dircetP}
\end{align}}
By multiplying inequality (\ref{eq:dircetP}) from both sides with $P^{-1}$ and using $F=GY$ and $Y=P^{-1}$, and finally applying the Schur complement, the inequality (\ref{eq:dircetP}) is transformed into 
\begin{align}
\begin{bmatrix}
        -Y & Y & (U_0 F)^T & (X_1 F)^T & F^T \\
        * & -Q^{-1} & 0 & 0 & 0\\
        * & * & -R^{-1} & 0 & 0 \\
        * & * & * & -\frac{Y}{\gamma} & 0 \\
        * & * & * & * & -\frac{1}{ \gamma Tr(PW)}I
        \end{bmatrix} \preceq 0. \label{eq:ddddSimple}
\end{align}
Under Assumptions 1-4, the solution exists if the problem is feasible. If there exists an $\alpha > 0$ such that 
\begin{align}
    Tr(PW) \le \frac{1}{\alpha}, \label{eq:PWa}
\end{align}
we can transform (\ref{eq:ddddSimple}) into (\ref{eq:lmiDD}). Using (\ref{eq:PYineq}), the inequality (\ref{eq:PWa}) can be written as 
\begin{align}
    \frac{1}{Tr(W^{-1} Y)} \le \frac{1}{\alpha n^2}, 
\end{align}
which can be translated into (\ref{eq:tracePW}). Also, instead of minimizing $Tr(PW)$, we minimize its upper bound by maximizing $\alpha.$ 

{
To find a lower bound for $\gamma,$ we need to demonstrate that the stability \eqref{eq:dircetP} of the discounted system implies the stability of the original system \eqref{eq:closedLoopDirect} using Lemma \ref{lemma:gamma-stability}
\begin{align}
    (X_1 G)^\top P (X_1 G) + Tr(PW)G^\top G  \preceq P.
\end{align}
Using \eqref{eq:dircetP}, we have
\begin{align}
    (1-\gamma)((X_1 G)^\top P (X_1 G) + Tr(PW)G^\top G) \nonumber \\ 
    \preceq Q + (U_0 G)^\top R (U_0 G),
\end{align}
which means that 
\begin{align}
    (1-\gamma)\lambda_{max}((X_1 G)^\top P (X_1 G) + Tr(PW)G^\top G) \nonumber  \\
     \preceq \lambda_{min}(Q + (U_0 G)^\top R (U_0 G)), 
\end{align}
which indeed is equivalent to \eqref{eq:data-gamma-lower}.
}

The LMI condition (\ref{eq:lmiDD}) is equivalent to the Lyapunov conditions in (\ref{eq:ddbell}). Based on Lemma 2 and Proposition 3 in \cite{pmlr-v120-coppens20a},
the direct data-driven closed-loop system (\ref{eq:multiSys}) is MSS.\qed
\end{theorem}
\begin{remark}
The approach presented to find the direct data-driven control gain is different from the method proposed in~\cite{DEPERSIS2021109548}. Leveraging the Lyapunov inequality (\ref{eq:ddbell}) to obtain a data-based formulation of the system (\ref{eq:multiSys}) directly leads to the data-driven Lyapunov inequality (\ref{eq:DDDD}). This enables us to take the covariance of the data-driven closed-loop system $\gamma Tr(PW)G^TG$ into account. Therefore, the resulting controller guarantees stability from the ground up. On the other hand, the method in~\cite{DEPERSIS2021109548} uses an SDP formulation of a model-based LQR problem that lacks consideration of the effect of noise on the closed-loop system. As a result, a posterior analysis is required to find the conditions of stability. The term $\gamma Tr(PW)G^T G$ is different from the term identified by~\cite{DEPERSIS2021109548} and is associated with the robustness of the direct data-driven LQR problem. In our derivation, the robustness term is weighted by the noise covariance $W$ and the value matrix $P$. This term has automatically emerged from the stability analysis and is not a regularization of the LQR cost. An interesting point is that the stochastic term $\Omega_0$ in closed-loop dynamics is taken into account during the controller design. The resulting direct data-driven LQR gain stabilizes the multiplicative noise system (\ref{eq:closedLoopDirect}) and the unknown dynamic system (\ref{eq:syst}) in the MSS sense.
\end{remark}
\textcolor{blue}{
\begin{remark}
    Even though Condition \eqref{eq:data-gamma-lower} shows a posterior-type of stability condition, it is common practice in discounted optimal control problems
 (see \cite{LAI2023110685,7588063,10251494,GAITSGORY2018311,9063479,LAI2023110685} for example). One can sidestep checking this posterior condition by picking a discount factor close to one to ensure that condition \eqref{eq:data-gamma-lower} is satisfied. Given a discount factor that satisfies the condition \eqref{eq:data-gamma-lower}, the presented data-driven approach provides an optimal robustness-performance trade-off \textit{by construction}, which achieves robustness by design in the presence of uncertainties. 
\end{remark}
}
\section{Suboptimality gap}
The analysis of the suboptimality gap between model-based and model-free LQR approaches has been done in the literature~\cite{tu2019gap,DEPERSIS2021109548}.
The suboptimality gap between the direct data-driven LQR (\ref{eq:directRobustlmi}) and the model-based LQR in Lemma 2 can be formalized as in Theorem 2. The following definitions will be employed in deriving the suboptimaltiy gap.
\textcolor{blue}{
\begin{definition}
    Signal to noise ratio ($\textnormal{SNR}$) relation is defined as~\cite{10061542,9992770}
\begin{align}
    \textnormal{SNR} = \frac{1}{\norm{\Omega_0}\norm{D_0^{\dag}}}.
\end{align}
\end{definition}
\begin{definition} \cite{NEURIPS2019_5dbc8390}
    The growth or decay of powers of a square matrix $L=A+BK^*$ can be defined as 
\begin{align}
    \tau(\sqrt{\gamma} L, \rho) \coloneqq \sup\left\{\norm{\gamma^{k/2} L^k}\rho^{-k}: k\geq 0 \right\},
\end{align}
which is the smallest value such that $\norm{\gamma^{k/2} L^k} \leq \tau(\sqrt{\gamma} L, \rho) \rho^k$ for some positive $\rho < 1$. Also consider $\bar{\tau}=\frac{\tau(\sqrt{\gamma} L, \rho)}{1-\rho^2}$.
\end{definition}
}
\begin{theorem}
    Consider the model-based and direct data-based LQR problems in (\ref{eq:optlmiModel}) and (\ref{eq:directRobustlmi}), respectively. The suboptimality gap between the direct data-driven LQR cost $J$, and the model-based LQR cost $J^*$ can be upper bounded by
{
    \begin{align}
    J-J^* &\leq x_0^T (P-P^*) x_0 + \frac{\gamma}{1-\gamma}Tr\Bigl((P-P^* )W\Bigr) \nonumber \\
        & \leq \norm{\Delta P} (\norm{x_0}^2 + \frac{\gamma}{1-\gamma}\abs{Tr(W)}), \label{eq:Jineq}
\end{align}
}
where $\Delta P = P-P^*$ and, 
\begin{align}
    \norm{\Delta P} \leq \delta = \frac{\delta_1 }{\delta_2}, \label{eq:DeltaP}
\end{align}
with $\delta_1 > 0$ and $\delta_2 > 0$ defined as
\vspace{-20pt}
\begin{dmath}
    \delta_1 = \bar{\tau}\norm{X_0^{\dag}+\bar{G}}^2 \left( 
    \norm{U_0}^2 \norm{R} + 2\gamma \norm{D_0}^2 \norm{P^*}  \left[\norm{\Theta}^2+\frac{1}{\textnormal{SNR}^2}\right] + \gamma \norm{P^*} Tr(W)\right)
    + \bar{\tau}\gamma \norm{A}\norm{P^*} \norm{L}, \label{eq:delta1}
\end{dmath}
\vspace{-20pt}
and
\begin{align}
    \delta_2 = 1-d,\label{eq:delta2} 
\end{align}
\vspace{-20pt}
\begin{dmath}
    d  = \bar{\tau}\gamma \norm{X_0^{\dag}+\bar{G}}^2 \left( 
    2\norm{D_0}^2 \left[\norm{\Theta}^2 + \frac{1}{\textnormal{SNR}^2}\right] + Tr(W)
    \right) + \bar{\tau}\gamma \norm{L}^2. \label{eq:c}
\end{dmath}
\vspace{-20pt}
Also, $\bar{G}$ shows the portion of the gain $G$ that belongs to the null space of $X_0$, and 
$\Theta=\begin{bmatrix}
    A & B
\end{bmatrix}$.\\
\textbf{Proof.} 
The inequality in (\ref{eq:Jineq}) is achieved using the trace-spectral norm inequality in~\cite{gravell2019learning}. To compute the suboptimality gap, we need to find an upper bound for $\Delta P$. The following analysis is inspired by~\cite{NEURIPS2019_5dbc8390}. Consider the following matrix expressions
\begin{align}
    &M(X) = X-\gamma A^T X A  - Q \nonumber \\
    & + \gamma^2 A^T X B (R + \gamma B^T X B)^{-1} B^T X A  \nonumber \\
    &= X-Q-\gamma A^T X (I+\gamma E X)^{-1}A,  \label{eq:M}\\
    &N(Z) = Z-Q \nonumber \\
    &+(U_0 X_0^{\dag} + U_0 \bar{G})^T R (U_0 X_0^{\dag} + U_0 \bar{G})\nonumber \\
    &-\gamma(X_1 X_0^{\dag}+X_1 \bar{G})^T Z (X_1 X_0^{\dag}+X_1 \bar{G}) \nonumber \\
    & -\gamma Tr(Z W) (X_0^{\dag} + \bar{G})^T (X_0^{\dag} + \bar{G}), \label{eq:N}
\end{align}
where $E=BR^{-1}B^T$. It is evident that $M(P^*)=0$ and $N(P) = 0.$ In (\ref{eq:N}), we used the fact that $G = X_0^{\dag} + \bar{G}$, where $\bar{G}$ belongs to the null space of $X_0$. The idea is to construct an operator $\Phi$ using $M$ and $N$ such that $\Delta P$ is its unique fixed point solution. Then, we find an upper bound for this solution based on available data and noise characteristics. It is followed from~\cite{NEURIPS2019_5dbc8390} that
\begin{align}
    M(&P^*+X) = T(X) + H(X), \label{eq:MPX}
\end{align}
where
\begin{align}
    T(X)&=X - \gamma L^T X L, \nonumber \\
    H(X) &= \gamma (A^T P^* +L^T X)L \nonumber \\
    &-\gamma A^T (P^* + X) (I+E(P^*+X))^{-1} A, \nonumber
\end{align}
where $L=A+BK^*$. The equation (\ref{eq:MPX}) is valid for any $X$ if the matrix $I+E(P^*+X)$ is invertible. It is also evident that by definition, $N(P)=N(P^*+X)=0.$ Now, consider the following matrix equation
\textcolor{blue}{
\begin{align}
M(P^*+X)-N(P^*+X) =  T(X) + H(X). \label{eq:MPXNPX}
\end{align}
}
It has a unique symmetric solution $X=\Delta P$ such that $P^*+X \succeq 0.$ As $L$ is a stable matrix, the linear map $T(X)$ is invertible~\cite{NEURIPS2019_5dbc8390} and its inverse (which is also a linear operator) can be computed as 
\begin{align}
\Phi(X)=T^{-1}\big(M(P^*+X)-N(P^*+X)-H(X)\big), \label{eq:Phi}
\end{align}
where $X=\Phi(X)$ and again $X=\Delta P$ is the unique symmetric fixed point solution to $\Phi.$ So finding $X$ using (\ref{eq:MPXNPX}) is equivalent to solving for $X$ using (\ref{eq:Phi}). Now, consider the following set
\begin{align}
    S_\delta = \left\{ X: \norm{X} \leq \delta, X=X^T, P^*+X \succeq 0 \right\}.
\end{align}
If we can find an upper bound $\delta$ based on available information and SNR ratio such that the operator $\Phi$ can map $S_\delta$ into itself and also be contractive over this set, then $\Phi$ has a fixed point in the set. Since $\Delta P$ is the only solution for $\Phi$, then we can conclude that $\norm{\Delta P} \leq \delta.$
At this point, we must find an upper bound for (\ref{eq:Phi})
\begin{align}
     &\norm{\Phi(X)} \nonumber \\
     &\leq \norm{T^{-1}(M(P^*+X)-N(P^*+X)-H(X))}. \label{eq:NormPhi}
\end{align}
This inequality holds as $T^{-1}$ is a linear operator. Since $L$ is a stable matrix, $X-\gamma L^T X L= S$ has a solution of the form
\begin{align}
    X=\sum_{k=0}^{\infty}{\gamma^k (L^k)^T S L^k}, \label{eq:Tinv}
\end{align}
for some matrix $S.$ It is clear from (\ref{eq:Tinv}) that 
\begin{align}
    \norm{X}=\norm{T^{-1}(S)}\leq \bar{\tau} \norm{S}, \label{eq:NormTinv}
\end{align}
By comparing (\ref{eq:NormPhi}) and (\ref{eq:NormTinv}), we have 
\begin{align}
    \norm{X} &\leq \frac{\tau(\sqrt{\gamma} L, \rho)}{1-\rho^2} \norm{S}, \label{eq:Xineq}
\end{align}
with $S=M(P^*+X)-N(P^*+X)-H(X).$ Using (\ref{eq:M}) and (\ref{eq:N}), $S$ can be simplified as
\begin{align}
   S &=  -(U_0 X_0^{\dag} + U_0 \bar{G})^T R (U_0 X_0^{\dag} + U_0 \bar{G})\nonumber \\
    &+\gamma(X_1 X_0^{\dag}+X_1 \bar{G})^T (P^*+X) (X_1 X_0^{\dag}+X_1 \bar{G}) \nonumber \\
    & +\gamma Tr((P^*+X) W) (X_0^{\dag} + \bar{G})^T (X_0^{\dag} + \bar{G}) \nonumber \\
    &-\gamma (A^T P^* +L^T X)L. \label{eq:Sequation}
\end{align}
Using the trace-spectral norm inequality, the norm of $S$ can be written as
\begin{align}
    \norm{S} &\leq \norm{U_0}^2 \norm{X_0^{\dag}+\bar{G}}^2 \norm{R} \nonumber \\
    &+ \gamma \norm{X_1}^2 \norm{X_0^{\dag}+\bar{G}}^2 \norm{P^*+X} \nonumber \\
    &+ \gamma \norm{X_0^{\dag}+\bar{G}}^2 \norm{P^*+X}Tr(W) \nonumber \\
    &+ \gamma \norm{L}^2 \norm{X} + \gamma \norm{A}\norm{P^*}\norm{L}. \label{eq:Sineq}
\end{align}
Also, considering the fact that $X_1 = A X_0 + B U_0 + \Omega_0$, and using the inequality $(a+b)^2<2(a^2+b^2)$, the $\norm{X_1}^2$ term can be upper bounded as
\begin{align}
    \norm{X_1}^2 \leq 2 \norm{D_0}^2 \left( \norm{\Theta}^2 + \frac{1}{\textnormal{SNR}^2} \right), \label{eq:X1SNR}
\end{align}
where we used the SNR relation and the inverse of spectral norm inequality~\cite{gravell2019learning}. The inequality (\ref{eq:X1SNR}) shows that the bound on $X_1$  increases if the SNR decreases. Now, using the fact that $\norm{P^*+X}\leq \norm{P^*}+\norm{X}$ and substituting (\ref{eq:Sineq}) and (\ref{eq:X1SNR}) in (\ref{eq:Xineq}), then $\norm{X}$ can be upper bounded by $\norm{X} \leq \delta = \frac{\delta_1 }{\delta_2}$.  
\textcolor{blue}{
Now, we only need to prove that the operator $\Phi$ is a contraction map for $X$ and $\hat{X}$ in $S_{\delta}$. Using (\ref{eq:NormTinv}), one has
\begin{align}
    \Phi(X)-\Phi(\hat{X}) = \sum_{k=0}^{\infty}{\gamma^k (L^k)^T (S(X)-S(\hat{X}))L^k},
\end{align}
where
\begin{align}
S(X)-&S(\hat{X})   \nonumber \\
 &=\gamma(X_1 X_0^{\dag}+X_1 \bar{G})^T (X-\hat{X}) (X_1 X_0^{\dag}+X_1 \bar{G}) \nonumber \\
    & +\gamma Tr((X-\hat{X}) W) (X_0^{\dag} + \bar{G})^T (X_0^{\dag} + \bar{G}) \nonumber \\
    &-\gamma L^T (X-\hat{X})L.
\end{align}
If the following condition holds
\begin{align}
    &\norm{\Phi(X)-\Phi(\hat{X})} \nonumber \\
    &\leq \bar{\tau}\norm{S(X)-S(\hat{X})} \nonumber \\
    & \leq d\norm{X-\hat{X}}, \quad 0< d < 1.
\end{align}
where $d$ is defined in (\ref{eq:c}), then the existence and uniqueness of the solution $\Delta P$ to $\Phi(X)$ is ensured. \qed} 
\end{theorem} 
\begin{remark} According to $d<1$, if SNR is low, then $\norm{\bar{G}}$ must be low, and if SNR is high, $\norm{\bar{G}}$ can be high. Also, based on (\ref{eq:delta1}) and (\ref{eq:delta2}), it is clear that the discount factor $\gamma$ has a direct effect on $\Delta P$, in the sense that if we choose a small value for $\gamma$, the suboptimality gap becomes small. However, choosing a small value for $\gamma$ deteriorates the stability of the closed-loop system. 
\end{remark}
{
\begin{remark} Based on \eqref{eq:Jineq} and \eqref{eq:DeltaP}, the upper bound for the suboptimality gap between the presented model-free LQR and the model-based LQR  depends on $\gamma$, $\norm{\bar{G}}$, SNR, and $\norm{U_0}$. In fact, $\Delta P$ has a direct relation with $\gamma$, $\norm{\bar{G}}$, and $\norm{U_0}$, and also it has an inverse relation with SNR. The discount factor $\gamma$ is fixed and cannot be considered as a design parameter since it may deteriorate the stability. $\norm{U_0}$ cannot be decreased too much, as it may not excite the system well. As expected, the suboptimality gap shrinks as the SNR increases. Besides, it can also be decreased by reducing $\norm{\bar{G}}$. Note that the suboptimality gap does not converge to zero. This is the cost we pay for robustness, as the cost is optimized over a set of system models and not a single model. This is similar to the results of \cite{DEPERSIS2021109548} (please see their equation (28)).
\end{remark}
\begin{remark}
    The reduction of conservatism depends on several factors. As demonstrated by our theoretical analysis, one significant factor affecting the performance gap is the SNR, and, consequently, the data quality.
    Future research directions include the development of a selective attention or active learning mechanism to focus on or generate data samples that improve the performance gap and robustness as data streams online. Finding criteria for the selection or generation of informative samples has been a daunting challenge. One promising approach will be to incorporate prior uncertain physical knowledge into our learning framework. The criteria for selection and generation of high-quality data will be to reduce the size of the set formed by the intersection of the physics-informed set of models (sets conformed with prior knowledge) and the closed-loop data-based set of models (closed-loop models conformed with data). The robust LQR optimization is performed over the intersection set and reducing its size reduces the uncertainty level and thus the suboptimality gap. 
    Preliminary results in~\cite{10318172} for indirect learning have shown that this integration can reduce the conservatism of LQR. However, incorporating physical knowledge in direct data-driven approaches is more challenging and requires new advancements.
\end{remark}
}
\section{Monte Carlo Simulations}
{
            This section compares the presented direct data-driven LQR (\ref{eq:directRobustlmi}) with the model-based LQR (\ref{eq:optlmiModel}), the direct data-driven approach (low complexity LQR (LCLQR)) presented in~\cite{DEPERSIS2021109548}, the CE-promoting direct data-driven LQR (PCELQR) in~\cite{10061542}, the fully regularized LQR (FRLQR) which combines the regularizers of LCLQR and PCELQR,  the indirect data-driven robust LQR (RLQR)~\cite{treven2021learning}, and the indirect LQR method in \eqref{eq:celmi}. Specifically, we compare their costs, number of failures ($N_f$), and regularization parameters against different SNRs and various $Q$ and $R$ matrices. 
}
We used CVXPY~\cite{diamond2016cvxpy,agrawal2018rewriting} for modeling the convex optimization problems and MOSEK~\cite{mosek} as the solver\footnote{\url{https://github.com/raminux/python-control/tree/master/published_papers_repo/Automatica}
}.
\textcolor{blue}{The tuning parameters used in each of the methods are as follows: $\alpha>0$ is the robustness parameter in LCLQR, $\lambda>0$ is a regularization term in PCELQR, $\lambda>0$, and robustness $\alpha>0$ terms are regularization terms in FRLQR, and  $c_\delta$ is the $(1-\delta)$-quantile of the $\chi^2$ distribution with $n(n+m)$ degrees of freedom in RLQR. The RLQR operates on a set of identified models using data. To identify these sets of systems, this method applies a sequence of random signals to the system and finds high-confidence sets for the system matrices using the collected data. This is also explained in the Arxiv version of the RLQR paper~\cite{Treven2020LearningCF}. To do a fair comparison, all the methods are assumed to have access to the same data set.}

Consider the following two degrees of freedom quarter-car suspension system~\cite{rajamani2011vehicle,roh1999stochastic}
\begin{align}
    \dot{x}(t)&=\begin{bmatrix}
        0 &1 &0 & -1 \\
        -\frac{k_s}{m} & -\frac{b_s}{m_s} & 0 & \frac{b_s}{m_s} \\
        0 & 0 &0& 1\\
        \frac{k_s}{m_u} & \frac{b_s}{m_u} & -\frac{k_t}{m_u} & -\frac{b_s}{m_u}
    \end{bmatrix} x(t) \nonumber \\
    & \quad + \begin{bmatrix}
        0\\ \frac{1}{m_s} \\ 0 \\ -\frac{1}{m_u}
    \end{bmatrix} u(t) + \omega(t),
\end{align}
where $x=\begin{bmatrix}
    x_1 & x_2 & x_3 & x_4
\end{bmatrix}^T$. The states $x_1$, $x_2$, $x_3$, and $x_4$ show the suspension deflection, absolute velocity of sprung mass, tire deflection, and absolute velocity of unsprung mass, respectively. Also, $m_s=240~kg$, $m_u=36~kg$, $b_s=980~\frac{N.s}{m}$, $k_s=16000~\frac{N}{m}$, and $k_t=160000~\frac{N}{m}$ represent the quarter-car equivalent of vehicle body mass, the unsprung mass due to axle and tire, suspension damper, suspension spring, and vertical stiffness spring respectively. The sampling time used for data collection is $T_s=0.01~s.$ The noise vector $\omega(t)$ shows uncertainties in system dynamics due to modeling errors and random vertical road displacement, which is assumed to have a Gaussian distribution with zero mean and the following covariance matrix
\begin{equation*}
    W = \begin{bmatrix}
        0.0001 & 0.0& 0.0 & 0.0 \\
    0.0 & 0.00001 & 0.0 & 0.0 \\
    0.0 & 0.0 & r & 0.0 \\
    0.0 & 0.0 & 0.0 & 0.001
    \end{bmatrix},
\end{equation*}
where $r > 0$ shows the random road displacement variance. In simulations, we change $r$ (and, as a result, the SNR) to evaluate the performance of different learning algorithms under various SNRs. The control input $u(t)$ represents the active force actuator of the suspension system. {The following input is applied to the system
\begin{align}
    u_k = 10\upsilon_k,
\end{align}
where $\upsilon_k$ is a zero mean and unit variance Gaussian noise.} We have generated $100$ sets of input-state data with $N=10$ and designed $N_K=100$ controllers using this data. For each of the designed controllers, $N_S=200$ simulations of $N_P=150$ sample points are performed. The initial conditions are drawn from a Gaussian distribution with mean $\begin{bmatrix}
    0.3 & -4 & 0.1 & -1
\end{bmatrix}^T$ and covariance of $0.0006 I$. 
{
\textcolor{blue}{Tables 1 and 2 summarize the simulation results for two sets of $Q$ and $R$ matrices.} The averaged cost of successful controller design \textcolor{blue}{$\bar{N}_K=N_K-N_f$} is computed using 
\begin{dmath}
    \bar{J}= \frac{1}{\bar{N}_K N_S N_P}\sum_{i=1}^{\bar{N}_K}{\sum_{j=1}^{N_S}{
    \sum_{k=0}^{N_P}{x_{i, j, k}^\top (Q+K_i^\top R K_i) x_{i,j, k}}
    }},
\end{dmath}
where $x_{i, j, k}$ represents the state vector at time $k$ in the $j-th$ simulation for the $i-th$ controller.\\
Table. 1 demonstrates results for the case where $Q$ is large compared to $R$ to favor stable designs. For SNR between $50-23 db$, the failure rate $N_f$ of LCLQR and PCELQR is the highest among the compared methods. In this method, \textcolor{blue}{ hyperparameters are chosen as $\alpha=10$ and $\lambda=0.01$ and are selected by trial and error to achieve the best results. In sharp contrast to our presented approach that makes a trade-off between robustness and performance \textit{by construction}, one of the main drawbacks of LCLQR and PCELQR is that higher-level optimization processes might be needed to learn the optimal set of hyperparameters.  Our approach in Theorem 1 designs the best controllers in this range in the sense of low $\bar{J}$ and $N_f$ for $\gamma=0.9999$. For the low SNR regime of $10 db$, the FRLQR with $\alpha=10$ and $\lambda=10$ has the lowest $N_f$ among the direct data-driven methods. Again, there is no systematic way to tune the hyperparameters using the FRLQR method.}

\textcolor{blue}{In Table 1, one can notice that the cost of model-based LQR is reasonably lower than others. This discrepancy has two main roots. First, in model-based LQR, the perfect knowledge of $A$ and $B$ is assumed to be available. Therefore, the exact LQR optimization is performed based on known system knowledge. However, in all other approaches for which $A$ and $B$ are learned, a single-point estimation of these matrices is not viable, and a set of possible models or a distribution of models is learned to explain the data. The cost is then robustly optimized over these sets. Second, the value of $Q$ is large, magnifying small differences in the system states.}

Table. 2 outlines the results for equal weights for states and input. \textcolor{blue}{In this case, the  LCLQR ($\alpha=0.01$),  PCELQR ($\lambda=1$), and FRLQR ($\alpha=0.01, \lambda=10$) methods result in acceptable average cost and small failures among various values for their hyperparamters. Note that in the existing data-based methods, comparing Tables 1 and 2, a different set of hyperparameters should be chosen for different sets of $Q$ and $R$ performance matrices. 
A main advantage of Theorem 1 is that by choosing $\gamma$ close to 1, there is no other hyper-parameter to tune the stability and performance dilemma. This is because our approach is robust by construction. This is a vital advantage as there is no specific approach in \cite{DEPERSIS2021109548,10061542} to hand-tune these hyperparameters to ensure robust performance.}

}
\textcolor{blue}{
\begin{remark}
    The best values of the parameters of methods in \cite{DEPERSIS2021109548} and ~\cite{10061542} are not known beforehand and depend on the cost function parameters. Therefore, finding the best set of hyper-parameters requires a higher-level optimization to optimize the trade-off between optimality and robustness, as pointed out in ~\cite{10061542}. Also, formalizing a higher-level cost for a higher-level optimization to find the best values of hyper-parameters is hard. This is because while some parameters improve the robustness, they ruin the performance, and vice versa. A significant advantage of our approach is that it bypasses fine-tuning of design parameters to achieve the best possible trade-off. That is, we provide a unique optimization that achieves optimal trade-off \textit{by construction} and does not require fine-tuning.
\end{remark}
}

\begin{table*}[]
  \caption{{Comparing different controllers in terms of their average costs $\bar{J}$ and number of failures ($Q=diag(\begin{bmatrix}
    30000, 30, 20, 1
\end{bmatrix}), R=0.0001$)}}
  \label{table:results}
  \centering 
  \begin{threeparttable}
    \begin{tabular}{ccccc}
    Controllers  & \multicolumn{3}{c}{\textbf{SNRs (dB)}} \\
     \midrule\midrule
      & 50  & 37  & 23 & 10 \\
      \midrule
       & $\bar{J},  n_f$  & $\bar{J}, n_f$ & $\bar{J}, n_f$ & $\bar{J}, n_f$  \\
LCLQR~\cite{DEPERSIS2021109548}  & &  & &   \\
    $\begin{aligned} 
    &\alpha = 10 
    \end{aligned}$ & $\begin{aligned}
        & 879.92, 7\\ 
    \end{aligned}$ & $\begin{aligned}
        & 947.29,  2\\ 
    \end{aligned}$ & $\begin{aligned}
        & 2753.12,  1 \\ 
    \end{aligned}$ & $\begin{aligned}
        & 38559.87, 3\\ 
    \end{aligned}$ 
    \\
      \midrule
 PCELQR~\cite{10061542}  & & &  &   \\  
    $\begin{aligned}
    &\lambda = 0.01 
    \end{aligned}$ & $\begin{aligned}
        & 877.70,   2\\ 
    \end{aligned}$ & $\begin{aligned}
        & 951.32,   2\\  
    \end{aligned}$ & $\begin{aligned}
        &2775.75, 5  \\ 
    \end{aligned}$ & $\begin{aligned}
        & 38225.26, 27\\ 
    \end{aligned}$ 
    \\
    \midrule
 FRLQR~\cite{10061542}  & &  &  &   \\
    $\begin{aligned} 
    &\alpha = 10, \, \lambda=10 
    \end{aligned}$ & $\begin{aligned}
        & 878.65,  1\\ 
    \end{aligned}$ & $\begin{aligned}
        & 947.26,  0\\  
    \end{aligned}$ & $\begin{aligned}
        &2749.82, 0 \\ 
    \end{aligned}$ & $\begin{aligned}
        &38508.27 , 0  \\ 
    \end{aligned}$ 
    \\
  \midrule
 CE (Lemma 4)  &   &  & &   \\
    $\begin{aligned} 
    &\gamma = 0.9999 
    \end{aligned}$ & $\begin{aligned}
        &933.64, 1\\  
    \end{aligned}$ & $\begin{aligned}
        &999.60,  3  \\    
    \end{aligned}$ & $\begin{aligned}
        &2878.22, 3\\ 
    \end{aligned}$ & $\begin{aligned}
        &40108.40, 1  \\ 
    \end{aligned}$  
    \\
    \cmidrule(l  r ){1-5}
RLQR~\cite{treven2021learning} &  &  & &   \\
    $\begin{aligned}
        &c_{\delta}=7.43 
    \end{aligned}$ & $\begin{aligned}
        &865.12,   0\\  
    \end{aligned}$  & $\begin{aligned}
        &945.00,  0 \\ 
    \end{aligned}$  & $\begin{aligned}
        &2752.92, 0  \\   
    \end{aligned}$  & $\begin{aligned}
        &38209.19,  0\\  
    \end{aligned}$  
    \\
    \cmidrule(l r ){1-5}
Theorem 1 & $\bar{\alpha}, \bar{J}, n_f$  & $\bar{\alpha},\bar{J}, n_f$ & $\bar{\alpha},\bar{J}, n_f$ & $\bar{\alpha},\bar{J}, n_f$ \\ 
$\begin{aligned}
 &\gamma = 0.9999 
\end{aligned}$
     & $\begin{aligned}
        &139.49, 898.14, 0\\ 
    \end{aligned}$  & $\begin{aligned}
        &134.09, 943.14,0  \\ 
    \end{aligned}$  & $\begin{aligned}
        &136.54, 2807.32, 0 \\ 
    \end{aligned}$  & $\begin{aligned}
        &139.04, 38537.07, 7  \\ 
    \end{aligned}$  
\\
 \cmidrule(l r ){1-5}
Model-Based LQR & & & & \\
$\begin{aligned}
      \bar{J}
\end{aligned}$
    &$272.45$   & $277.97$ & $540.31$ & $6067.78$
\\
    \midrule\midrule
    \end{tabular}
\end{threeparttable}    
  \end{table*}

\begin{table*}[]
  \caption{{Comparing different controllers in terms of their average costs $\bar{J}$ and number of failures ($Q=I, R=1$)}}
  \label{table:results}
  \centering 
  \begin{threeparttable}
    \begin{tabular}{ccccc}
    Controllers  & \multicolumn{3}{c}{\textbf{SNRs (dB)}} \\
     \midrule\midrule
      & 50  & 37  & 23 & 10 \\
      \midrule
        & $\bar{J}, n_f$  & $\bar{J}, n_f$ & $\bar{J},  n_f$ & $\bar{J}, n_f$  \\
LCLQR~\cite{DEPERSIS2021109548}  & &  & &  \\
    $\begin{aligned} 
    &\alpha = 0.01
    \end{aligned}$ & $\begin{aligned}
        & 3.54,  1 \\  
    \end{aligned}$ & $\begin{aligned}
        & 5.088, 0 \\ 
    \end{aligned}$ & $\begin{aligned}
        & 44.16,  0 \\ 
    \end{aligned}$ & $\begin{aligned}
        & 817.23,  1 \\ 
    \end{aligned}$ 
    \\
      \midrule
 PCELQR~\cite{10061542}  &   &  & &   \\  
    $\begin{aligned} 
    &\lambda = 1 
    \end{aligned}$ & $\begin{aligned}
        & 3.55, 0 \\ 
    \end{aligned}$ & $\begin{aligned}
        & 5.083, 0 \\ 
    \end{aligned}$ & $\begin{aligned}
        & 44.09, 0 \\  
    \end{aligned}$ & $\begin{aligned}
        & 812.81, 0\\ 
    \end{aligned}$ 
    \\
    \midrule
 FRLQR~\cite{10061542}  &  &  &  &   \\
    $\begin{aligned} 
    &\alpha = 0.01,\, \lambda=10 
    \end{aligned}$ & $\begin{aligned}
        & 3.69, 0 \\  
    \end{aligned}$ & $\begin{aligned}
        & 5.16,  0 \\ 
    \end{aligned}$ & $\begin{aligned}
        & 44.51,  0 \\ 
    \end{aligned}$ & $\begin{aligned}
        & 822.20,  0 \\ 
    \end{aligned}$ 
    \\
  \midrule
 CE (Lemma 4)  &   & &  &   \\
    $\begin{aligned} 
    &\gamma = 0.9999 
    \end{aligned}$ & $\begin{aligned}
        & 7449.99,  0 \\ 
    \end{aligned}$ & $\begin{aligned}
        & 10915.21,1 \\ 
    \end{aligned}$ & $\begin{aligned}
        & 33693.9,  0 \\ 
    \end{aligned}$ & $\begin{aligned}
        & 335392.96, 0 \\ 
    \end{aligned}$  
    \\
    \cmidrule(l  r ){1-5}
RLQR\cite{treven2021learning} &   & & &   \\
    $\begin{aligned}
        &c_{\delta}=7.43 \\
    \end{aligned}$ & $\begin{aligned}
        & 3.56, 0 \\ 
    \end{aligned}$  & $\begin{aligned}
        & 5.13,  0 \\ 
    \end{aligned}$  & $\begin{aligned}
        & 44.17, 0 \\ 
    \end{aligned}$  & $\begin{aligned}
        & 814.64, 0 \\ 
    \end{aligned}$  
    \\
    \cmidrule(l r ){1-5}
Theorem 1 & $\bar{\alpha}, \bar{J},  n_f$  & $\bar{\alpha},\bar{J},  n_f$ & $\bar{\alpha},\bar{J}, n_f$ & $\bar{\alpha},\bar{J},  n_f$ \\ 
$\begin{aligned}
 &\gamma = 0.9999
\end{aligned}$
     & $\begin{aligned}
        & 3426.19,3.55,  0 \\ 
    \end{aligned}$  & $\begin{aligned}
        & 3344.39, 5.082,  0\\ 
    \end{aligned}$  & $\begin{aligned}
        & 3328.68, 44.038,  0 \\ 
    \end{aligned}$  & $\begin{aligned}
        & 3380.81, 814.25,  0 \\ 
    \end{aligned}$  
\\
 \cmidrule(l r ){1-5}
Model-Based LQR & & & & \\
$\begin{aligned}
      \bar{J}
\end{aligned}$
    &$3.52$   & $5.07$ & $44.02$ & $810.25$
\\
    \midrule\midrule
    \end{tabular}
\end{threeparttable}    
  \end{table*}

%
  


\section{Conclusion}
In summary, the paper introduces a one-shot direct data-driven controller for the infinite-horizon discounted LQR problem. Unlike traditional control methods, this approach relies on data without requiring a system identification step. Leveraging stochastic Lyapunov theory and Bellman inequality, the study formulates an SDP solution, ensuring robustness guarantees at the design stage without explicit regularization. The proposed approach accounts for the noisy characterization of the closed-loop system, eliminating the need to manually tune the trade-off between robustness and performance. The closed-loop system is revealed to have multiplicative and additive noises, offering opportunities to design distributionally robust direct data-driven LQR algorithms. Analysis of the suboptimality gap highlights the impact of SNR, discount factor, and the norm of input signals on controller performance. The simulation results on an active car suspension problem demonstrate the superiority of the proposed approach in terms of both robustness and performance gap compared to existing methods. By addressing limitations and providing a systematic approach for robust LQR design, this work contributes to the advancement of efficient data-driven control strategies. In future work, several avenues for further exploration and enhancement of the proposed one-shot direct data-driven LQR problem can be pursued, such as extending to nonlinear systems, real-world implementation, and addressing multiagent systems.

\begin{ack}                               
 This research is partially supported by the Department of Navy award N00014-22-1-2159, and partially supported by the National Science Foundation (NSF) award ECCS-2227311.  
\end{ack}

\bibliographystyle{plain}        
\bibliography{autosam}           

\end{document}